\documentclass[twoside,a4paper,11pt]{sea10}
\usepackage{graphicx}
\usepackage{hyperref}
\usepackage{movie15}
\begin{document}
\pagenumbering{arabic}
\pagestyle{myheadings}
\thispagestyle{empty}
{\flushleft\includegraphics[width=\textwidth,bb=58 650 590 680]{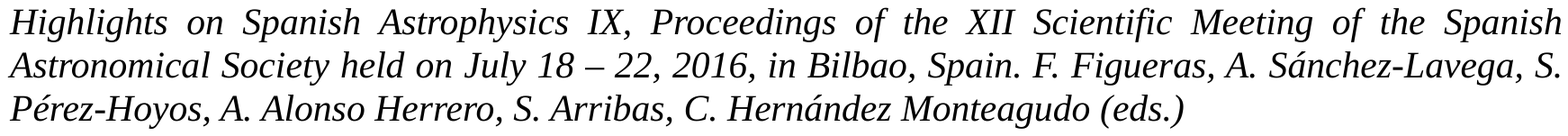}}
\vspace*{0.2cm}
\begin{flushleft}
{\bf {\LARGE
%
Music and astronomy. II.  $\bullet$unitedsoundsofcosmos
%
}\\
\vspace*{1cm}
%
J. A. Caballero$^{1,2}$,
A. Arias$^{3,4}$, 
J.~J. Machuca$^{3,5}$, 
and 
S. Morente$^{3,6}$
%
}\\
\vspace*{0.5cm}
%
$^{1}$ Landessternwarte K\"onigstuhl, ZAH, K\"onigstuhl 12, D-69117 Heidelberg, Germany \\
$^{2}$ Centro de Astrobiolog\'ia, CSIC-INTA, ESAC Camino Bajo del Castillo s/n, E-28692 Villanueva de la Ca\~nada, Madrid, Spain\\
$^{3}$ {\em Los Evangelistas}, Granada, Spain\\
$^{4}$ {\em Lagartija Nick}, Granada, Spain\\
$^{5}$ {\em Lori Meyers}, Granada, Spain\\
$^{6}$ {\em Sole\'a Morente}, Granada, Spain
%
\end{flushleft}
%
\markboth{
$\bullet$united{\bf sounds}of{\bf cosmos}
}{ 
%
Caballero et~al.
%
}
\thispagestyle{empty}
\vspace*{0.4cm}
\begin{minipage}[l]{0.09\textwidth}
\ 
\end{minipage}
\begin{minipage}[r]{0.9\textwidth}
\vspace{1cm}
\section*{Abstract}{\small
%
We have been congratulated on the stage by a Nobel laureate (he was our curtain raiser), played our music in planetariums, museums, observatories throughout Spain and at the end of the meeting of the ESO telescopes time allocation committee, shocked audiences in rock concerts, written monthly on {\em Musica Universalis}, made the second concert in 3D in Spain after Kraftwerk and broadcasted it live in Radio~3, mixed our music with poetry read aloud by scientists, composed the soundtracks of CARMENES, QUIJOTE, ESTRACK and the {\em Gaia} first data release, made a videoclip on how computer simulates the formation of stars... 
All those moments will not be lost in time like tears in rain, but put together in Bilbao during the 2016 meeting of the Spanish Astronomical Society.
%
\normalsize}
\end{minipage}
%
%
%
\section{\em O Fortuna velut luna statu variabilis... \label{intro}}

Why do you read this proceeding of the meeting of a national astronomical society? 
Probably because of the same reason as for me to write it: because we love music.
Do you remember when did your interest in space sciences start? And in music? 
The first author (JAC) is an astronomer because he watched {\em Stars Wars} and {\em The Empire Strikes Back} when he was four years old, and since then he has always wanted to ``explore'' other worlds, in~situ or, more realistically, from the ground with state-of-the-art instruments and telescopes of all sizes.
His interest in music came later, when he was already a teenager.
The first compact disc that he ever bought was {\em The Songs of Distant Earth} (1994), but he had already all the Mike Oldfield's discography in cassette.
The lyrics of one of the Oldfield's songs, \href{https://www.youtube.com/watch?v=gvzG9siO704}{``Saved by a Bell''} in his album {\em Discovery} (1984), read like this: \\

{\em
Would you like to look through my telescope? \\
\indent The Milky Way's a fine sight to see. \\
\indent All around our universe, we try so hard to view \\
\indent What's new.

Make a trip down to Sagittarius \\
\indent And take a spin by some nebula. \\
\indent I hope the sky stays clear for us, the night goes on so far \\
\indent In stars.

[...]

Shining like bright diamonds, the galaxies. \\
\indent Jupiter and Saturn spin by. \\
\indent Passing by companions, they all go drifting by. \\
\indent They fly!

Carry me down to see Aquarius. \\
\indent We're hoping to meet a shooting star. \\
\indent I can see there's going to be a message from afar. \\
\indent How close we are. \\
}

At one point, JAC was suggested to listen to {\em Di\'alogos Tres}, a daily programme on world music, ambient and new age in Radio~3 (the Spanish analogue to BBC Radio~1, 2 and 3, but with no classical music).
After being hooked on it, he~started listening the following programme in the radio dayparting, {\em El Ambig\'u}, and later the following one, and when he~started the grade of Physics he~did not know how to study without listening music of any style (Radio~3, Radio~Cl\'asica or his own cassettes and CDs). 
His passion for music was so intense that every chapter of his PhD thesis \cite{Ca10} started with a piece of lyrics, such as Claude Bertout had done it for his review on T~Tauri stars with the Leonard Cohen's poem ``Another Night With Telescope'' \cite{Be89}:  \\

{\em
I know the stars \\
\indent are wild as dust \\
\indent and wait for no man's discipline \\
\indent but as they wheel \\
\indent from sky to sky they rake \\
\indent our lives with pins of light. \\
}

In the meantime, Enrique Morente, Antonio Arias (AA) and a few great musicians in Spain challenged the flamenco with a breaking album, \href{https://play.spotify.com/album/3D9NJfydhCXchMlIuah63L}{\em Omega} (1996), which has been played in concerts all over the world (New York, Ciudad de M\'exico, Buenos Aires, Paris, Marseille, Cannes, Bastia, Antwerp and the whole Spain).
Afterwards, Enrique Morente went on mixing the most traditional flamenco roots with other influences, while AA, as the leader of the rock band Lagartija Nick, started exploring new concepts, sometimes with astronomical inspiration and even lyrics.
For example, in the homonymic album {\em Lagartija Nick} (1999), he composed songs on pulsars, ether, spheres traveling in space, a Moon base, selenography, the experience of astronauts in space, HAL~9000 and even light pollution (\href{https://www.youtube.com/watch?v=xeNDuBjBRm8}{``Azora~67''}):  \\

{\em
Demasiada luz, demasiada luz \hspace{\stretch{1}} Too much light, too much light \\
\indent La luz ensucia el cielo \hspace{\stretch{1}} Light messes the sky \\
\indent Mi cielo est\'a vac\'io con demasiada luz \hspace{\stretch{1}} My sky is empty with too much light \\
\indent La luz oculta estrellas  \hspace{\stretch{1}} Light hides stars \\
\indent Mi cielo est\'a vac\'io con demasiada luz \hspace{\stretch{1}} My sky is empty with too much light \\
}

In 2007, JAC published an outreach paper in Astronom\'iA, the Spanish counterpart of Sky \& Telescope or Sterne und Weltraum, on examples of musical astronomy and astronomical music \cite{Ca07}.
Just afterwards, the paths of JAC, a professional astrophysicist expert in stars, brown dwarfs, planets and instrumentation, and AA, a professional musician, composer, vocalist, bassist and guitarist, merged into a single astro-musical project, dubbed $\bullet$united{\bf sounds}of{\bf cosmos} \cite{Cab10, Ca12a, Ca12b}.

We are far from being pioneers in the use of music for education and outreach of astronomy.
For example, Carl Sagan et~al. \cite{Sa78} or Andrew Fraknoi \cite{Fr08, Fr15} already set up other comprehensive lists of astro-musical examples when some of the authors of this contribution had not been born yet.
However, our aim here is to show how we use 21st century tools for communicating ``astronomy for the masses'' (in Depeche Mode's~words).

\section{\em Ground Control to Major Tom}


\begin{figure}
\center
\includegraphics[height=0.46\textwidth]{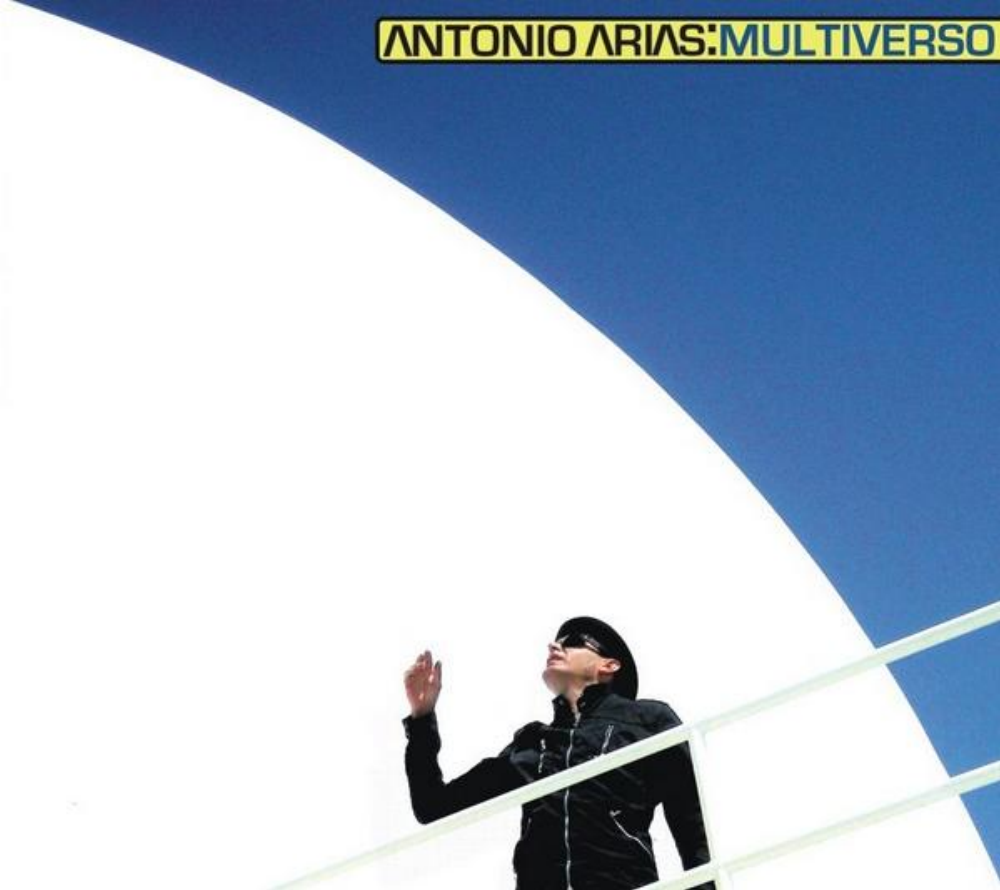}
\includegraphics[height=0.46\textwidth]{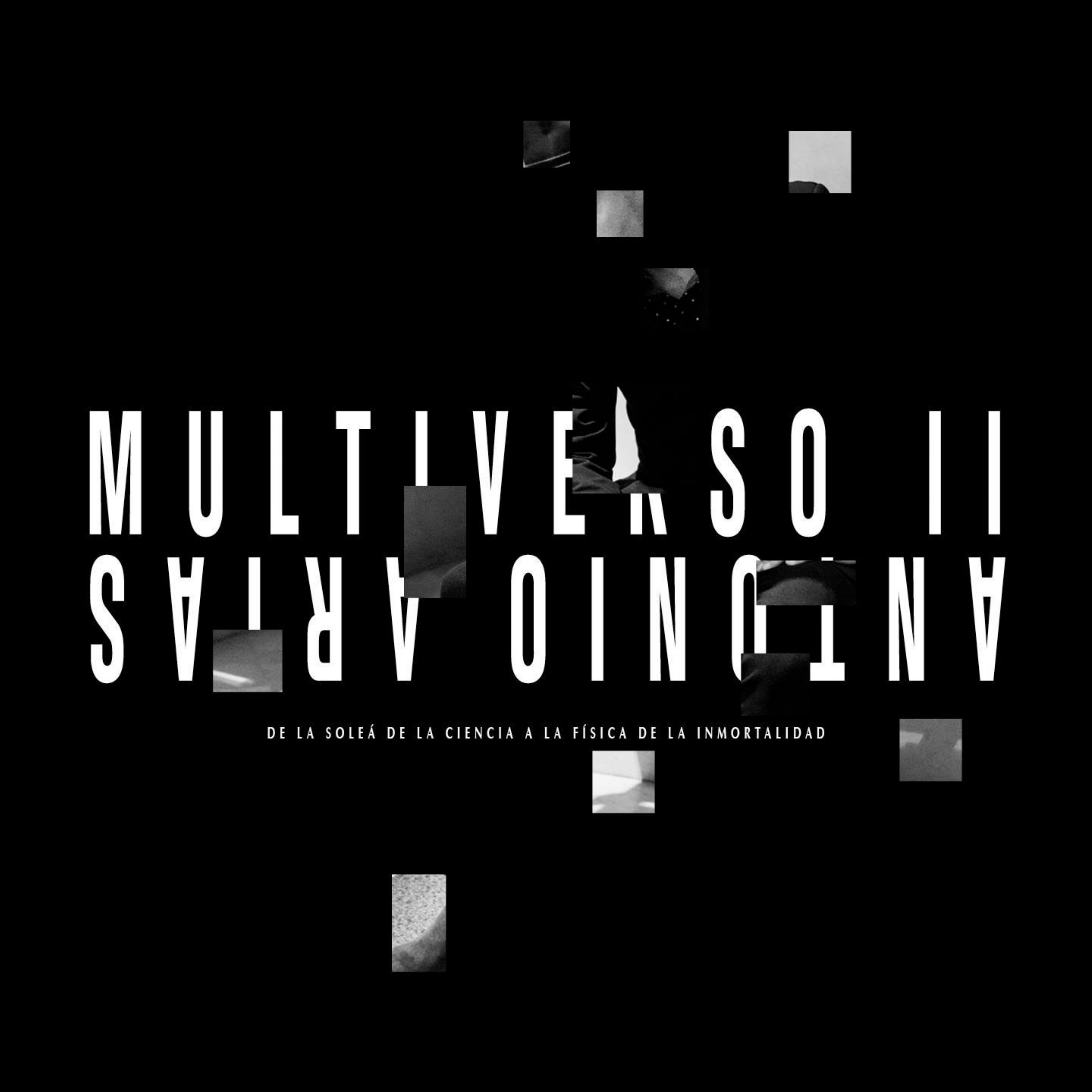}
\caption{\label{fig1} Covers of the Antonio Arias' albums {\em Multiverso} (2009, left, with the dome of the 2.2\,m Calar Alto telescope in the background) and {\em Multiverso II. De la sole\'a de la ciencia a la f\'isica de la inmortalidad} (2013, right).}
\end{figure}

\begin{figure}
\center
\includegraphics[width=0.99\textwidth]{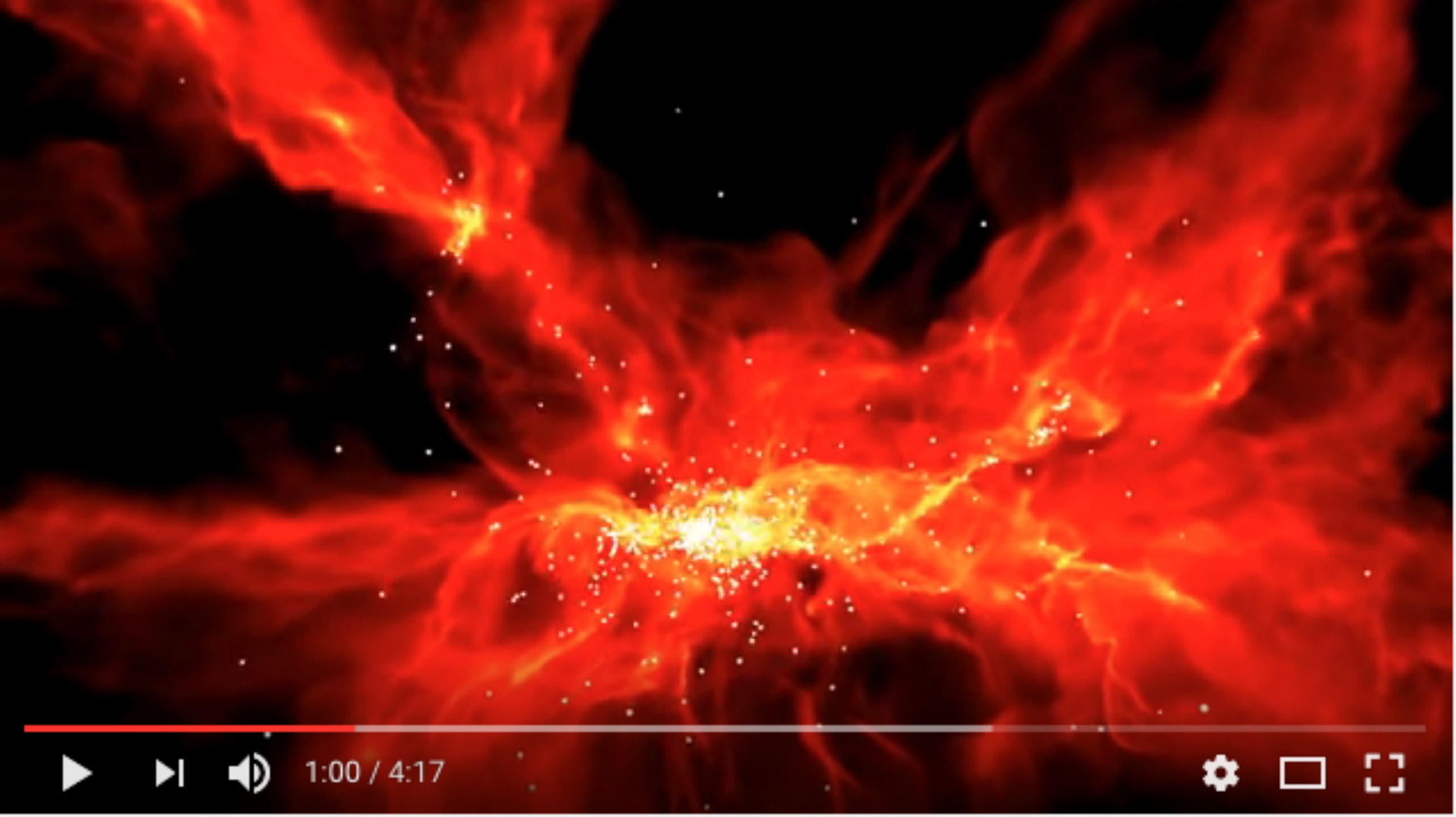}
\caption{\label{fig2} Caption of the videoclip of ``El ordenador simula el nacimiento de las estrellas'' ({\em Multiverso} 2009).
Music by Antonio Arias, lyrics by David Jou, original simulation by Matthew Bate, videoclip by David Cabezas and Jos\'e~A. Caballero, and special effects by David Callej\'on and Javier Fern\'andez.
Video available at \href{https://www.youtube.com/watch?v=J9lCSCV3Mkk}{YouTube}.}
\end{figure}

The origin of {\em Multiverso} (2009; Fig.~\ref{fig1}, left panel), the first AA's solo album outside Lagartija Nick and 091\footnote{Cero noventa y uno, 091, was a famous Spanish rock band active in the 80s and early 90s.} was another outreach paper in Astronom\'iA by JAC on poetry and astronomy \cite{Ca09}.
The title of the album mixed cosmological (``multiverses'') and poetical (``multi-verses'') concepts.
{\em Multiverso} started with the noises of the dome of the 2.2\,m Calar Alto telescope, continued with songs with lyrics from poems composed by Carlos Marzal, Natalia Carbajosa, Jos\'e Emilio Pachecho or, especially, David Jou, and finished with a 21st-century revision of the Johannes Kepler's {\em Harmonices Mundi} (Music of the Spheres).
The album was premiered during the closing ceremony of the International Year of Astronomy 2009 in Spain.   
The second track in {\em Multiverso} was ``El ordenador simula el nacimiento de las estrellas'' (Computer simulates the formation of stars; Fig.~\ref{fig2}), for which we produced a videoclip with real simulations by Bate et~al. \cite{Ba03}.
The other songs had titles such as ``Desde una estrella enana'' (on a G2\,V star with an old planet), ``G\'enesis'' (on the big bang) or ``La derrota de Bill Gates'' (on the effect of a coronal mass ejection on the Earth surface). 

{\em Multiverso II. De la sole\'a de la ciencia a la f\'isica de la inmortalidad} (2013; Fig.~\ref{fig1}, right panel) went on mixing scientific poetry with electric and bass guitars, drums and keyboards.
The last track of the album was the first of our soundtracks for astronomical instruments, facilities or data releases:
\href{https://www.youtube.com/watch?v=YT36FUHXq8c}{``C.A.R.M.E.N.E.S.''}. 
CARMENES is the new optical and near-infrared high-resolution spectrograph at the 3.5\,m Calar Alto, especially designed for the discovery of exoearths in the habitable zone around M dwarfs with the radial-velocity method \cite{Qu14}.
There was a version of the soundtrack in Spanish starred by Sole\'a Morente, the youngest Enrique Morente's daughter, and another \href{https://www.youtube.com/watch?v=sgV8yIZ-E9c}{in~English}, which was played live during a concert in one of the telescope domes of the Calar Alto observatory.

{\em Multiversos} (2015) was a box set compilation that gathered both {\em Multiverso} and {\em Multiverso~II} in vinyl LPs, together with four digital downloadable tracks, which are the seed for a future {\em Multiverso~III}. 
One of the four tracks was \href{https://www.youtube.com/watch?v=gRUH3i9ZCX4}{``Q-U-I JOint TEnerife''}.
QUIJOTE is a set of two telescopes and three instruments at the Observatorio del Teide that measure the polarisation of the cosmic microwave background radiation in the 11--40\,GHz frequency range with a spatial resolution of 1\,deg \cite{RM12}.  
Some members of $\bullet$united{\bf sounds}of{\bf cosmos} travelled to Tenerife, where staff of the Instituto de Astrof\'isica de Canarias filmed and edited the videoclip for the QUIJOTE soundtrack.

\href{https://soundcloud.com/user-91608032/villafrancacebrerosnew-norciamalargue-tambien-brilla-la-materia-1}{``Villafranca/Cebreros/New Norcia/Malarg\"ue (tambi\'en brilla la materia)"} (soundcloud, 2015) was our answer to the European Space Agency's ESTRACK 40th Anniversary Sound Contest. 
Our brand-new rock song begun with the names of the ESA ground-based space-tracking stations worldwide (some of which appear in the title) and followed with our characteristic astro-poetical lyrics in Spanish ({\em hidden behind immense clouds, matter also shines...}). 
We accompanied our music with, e.g., an Ariane~5 go/no-go pre-launch sequence, the Sputnik~I's beep-beep and the asteroseismological sounds of a pulsating star. 

The instrumental song \href{https://play.spotify.com/album/0pEmWqj3BhxgglzT5OmJ20}{{\em ``Gaia} DR1 (a soundtrack for the ESA billion star surveyor)''} (spotify, 2016) was premiered contemporaneously to the {\em Gaia} First Data Release \cite{Ga16}.
The track duration of 63\,s reflected the 63\,d precession period of the {\em Gaia}'s Lissajous orbit around the Sun-Earth Lagrangian point L$_2$.


Besides the albums, we have also performed live shows, dubbed astroconcerts, in which we mix rock, pop, electronica, stellar astrophysics, introduction to astronomy, science in general, poetry and video art in different proportions depending on the audience and facilities.
Between December 2009 and May 2016, we have played 12 astroconcerts in 
Munich (Municon, ``ESO OPC P98 get-together''),
La~Laguna (Aguere Cultural),
Almer\'ia (Calar Alto observatory, live streaming),
Barcelona (CosmoCaixa, together with Prof. David Jou), 
Madrid (Sala El Sol, twice; IX Scientific Meeting of the Spanish Astronomical Society~\cite{Cab11}; XIX Congreso Estatal de Astronom\'{\i}a), 
and Granada (Palacio de Congresos, twice: one together with Prof. Robert W. Wilson, the other with 3D glasses, cinema-like projection screen and broadcasted live by Radio~3; Parque de las Ciencias, together with Prof. Emilio Alfaro; Instituto de Astrof\'isica de Andaluc\'ia), 
A number of musicians have collaborated in our astroconcerts, including members of rock bands Los~Planetas, Lori~Meyers, Lagartija~Nick and P\'ajaro~Jack (see Acknowledgements).


Since 2013, JAC is the contributing editor of {\em Musica Universalis}, a section of the Astronom\'iA magazine.
There, he writes monthly on the music that did {\em not} go in the Golden Voyager Record; 
Muse (the rock band), MUSE (the VLT instrument) and the Muses (Euterpe and Urania); 
Franco Battiato ({\it Telescopi giganti per seguire le stelle}); 
T-shirts with the Mozart's {\em Eine kleine Nachtmusik} score and the S.~Jocelyn Bell Burnell's Little Green Men-1 (PC~1919) pulsar...


Now, music and astronomy reaches a much wider audience, since JAC has closed the loop and now collaborates with \href{http://www.rtve.es/alacarta/audios/longitud-de-onda/}{\em Longitud de Onda}, a Radio Cl\'asica programme. 
There, he talks every other week on the music that {\em did} go in the Golden Voyager Record; 
flamenco and astronom\'ia through the figures of {\em El Planeta}, the first reported flamenco artist, and Enrique Morente;
science-fiction films that happen beyond the Earth's low orbit and which soundtracks have been awarded or nominated to the Academy Award to the Best Score;
F.~William Herschel, who composed 24 symphonies and many concertos, apart from discovering Uranus, Titania, Oberon, Enceladus, Mimas and infrared radiation, and building the largest telescope of the world for 50 years... 

%

%
%
\small  
%
\section*{Acknowledgments}   
%
JAC is a Klaus Tschira Stiftung postdoctoral fellow at the LSW.
We thank the Sociedad Espa\~nola de Astronom\'ia for their support.
Other artists who have participated in our astroconcerts are:
Xarim~Arest\'e, Carmen~Arias, Juano~Azagra, Jaime~Beltr\'an, Juan\,B.~Codorn\'iu, David~Fern\'andez, Mafo~Fern\'andez, Nayra~Garc\'ia, Alfonso~Gonz\'alez (Popi), Daniel Guirado, Carlos~Gracia, Antonio~L\'opez (Noni), Migueline~L\'opez, Alejandro~M\'endez, Julian~M\'endez, Arturo~Mu\~noz, Florent~Mu\~noz, Juan~R.~Rodr\'iguez (J), Mario~Rodr\'iguez and Pepe~Ruiz.
%

%

\begin{thebibliography}{}
\small
%
\bibitem{Ba03}{Bate, M.~R. et~al. 2003, MNRAS, 339, 577}
\bibitem{Be89}{Bertout, C. 1989, ARA\&A, 27, 351}
\bibitem{Ca07}{Caballero, J. A. 2007, Astronom\'iA, 95, 26} 
\bibitem{Ca09}{Caballero, J. A. 2009, Astronom\'iA, 116, 24} 
\bibitem{Ca10}{Caballero, J. A. 2010, ASSP, 14, 74} 
\bibitem{Ca12a}{Caballero, J. A. 2012a, CAPJ, 12, 6} 
\bibitem{Ca12b}{Caballero, J. A. 2012b, Astronom\'iA, 161, 20} 
\bibitem{Cab10}{Caballero, J. A., Gonz\'alez~S\'anchez, S., Caballero, I. 2010, ASSP, 14, 548}
\bibitem{Cab11}{Caballero, J. A., Arias, A., Garc\'ia, N. 2011, hsa6, 857} 
\bibitem{Fr08}{Fraknoi, A. 2008, ASPC, 400, 514}
\bibitem{Fr15}{Fraknoi, A. 2015, ASPC, 500, 107}
\bibitem{Ga16}{Gaia Collaboration et~al. 2016, A\&A, in~press, eprint arXiv:1609.04172}
\bibitem{Qu14}{Quirrenbach, A. et~al. 2014, SPIE, 9147, E1F}
\bibitem{RM12}{Rubi\~no-Mart\'in, J. A. et~al. 2012, SPIE, 8444, E2Y}
\bibitem{Sa78}{Sagan, C. et~al. 1978, {\em Murmurs of Earth: The Voyager Interstellar Record}, ed. Random House, New York}
%
%
\end{thebibliography}
\end{document}